\documentclass[10pt]{article}

\usepackage{arxiv}
\usepackage{amsmath,amssymb}
\usepackage{booktabs}
\usepackage{graphicx}
\usepackage{multirow}
\usepackage{xcolor}
\usepackage{url}
\usepackage{float}

\title{AeroTSBoost: Temporal-Statistical Boosting for Real-World UAV Telemetry Anomaly Mining}

\author{
\begin{tabular}{c}
Junhao Wei\textsuperscript{1,2}, Haochen Li\textsuperscript{1}, Yanxiao Li\textsuperscript{1}, Yifu Zhao\textsuperscript{1}, Dexing Yao\textsuperscript{1}, Baili Lu\textsuperscript{1}\\
Xudong Ye\textsuperscript{1}, Sio-Kei Im\textsuperscript{3}, Yapeng Wang\textsuperscript{1}, and Xu Yang\textsuperscript{1,*}
\end{tabular}\\[0.5em]
\begin{tabular}{c}
\textsuperscript{1}Faculty of Applied Sciences, Macao Polytechnic University, Macao, 999078, China\\
\textsuperscript{2}Pazhou Lab (Huangpu), Guangzhou, 510555, China\\
\textsuperscript{3}Macao Polytechnic University, Macao, 999078, China\\
\textsuperscript{*}Corresponding author: Xu Yang (xuyang@mpu.edu.mo)
\end{tabular}
}
\date{}

\begin{document}
\maketitle

\begin{abstract}
Mining anomalies from unmanned aerial vehicle (UAV) state-estimation logs is challenging because failures are sparse, temporally structured, and distributed across heterogeneous PX4 telemetry streams with variable sensor availability and missing values. We present AeroTSBoost, a temporal-statistical boosting framework for real-world UAV telemetry anomaly mining. AeroTSBoost aligns multivariate flight logs, converts each window into deterministic descriptors that capture distributional shifts, quantile structure, endpoint drift, local dynamics, and lag correlation, and trains a class-balanced LightGBM detector. On UAV-SEAD, AeroTSBoost achieves the strongest AUPRC among evaluated classical, supervised tabular, neural reconstruction, recurrent, Granger-causality-based, and frequency-domain baselines. Across five seeds, it reaches $0.7516\pm0.0043$ AUPRC and $0.5342\pm0.0108$ threshold-swept event F1, improving AUPRC by 5.79 absolute points over the strongest non-AeroTSBoost baseline. Under purged chronological and leave-log-out protocols, it remains the best AUPRC method, reaching $0.6066\pm0.0193$ and $0.6388\pm0.0315$, respectively. On related ALFA fixed-wing UAV fault logs, AeroTSBoost reaches $0.9259\pm0.0076$ leave-sequence-out AUPRC, ahead of RandomForest ($0.8835\pm0.0797$) and moments-only ($0.8700\pm0.0481$). These results show that deterministic temporal-statistical representations remain highly competitive for sparse anomaly mining in operational cyber-physical telemetry.
\end{abstract}


\section{Introduction}
Unmanned aerial vehicles depend on state-estimation pipelines that fuse inertial, barometric, magnetic, visual, optical-flow, and global-positioning signals. Faults in this pipeline may appear as short transients, slow drifts, inconsistent cross-sensor relationships, or coupled changes across multiple telemetry streams. Detecting these failures from real flight logs is therefore a data mining problem with several difficult properties: the data are heterogeneous, temporal, high dimensional, sparsely labeled, and collected from cyber-physical systems operating under changing environmental and sensing conditions.

Recent multivariate time-series anomaly detectors have made significant progress through reconstruction, forecasting, graph learning, transformer attention, recurrent modeling, and frequency-domain modeling~\cite{deng2021gdn,xu2022anomalytransformer,tuli2022tranad,catch2025,xlstmad2025,gcad2025}. However, real UAV state-estimation logs expose practical challenges that are not always represented by generic public benchmarks. First, PX4 logs may have variable sensor availability and irregular sampling behavior. Second, anomalies occupy only a small fraction of windows, so AUROC alone can overstate detection quality while AUPRC and event-level metrics are more informative. Third, unsupervised reconstruction or forecasting losses do not necessarily optimize the sparse labeled event-detection objective that matters in operational telemetry mining.

\begin{figure}[t]
  \centering
  \includegraphics[width=0.98\textwidth]{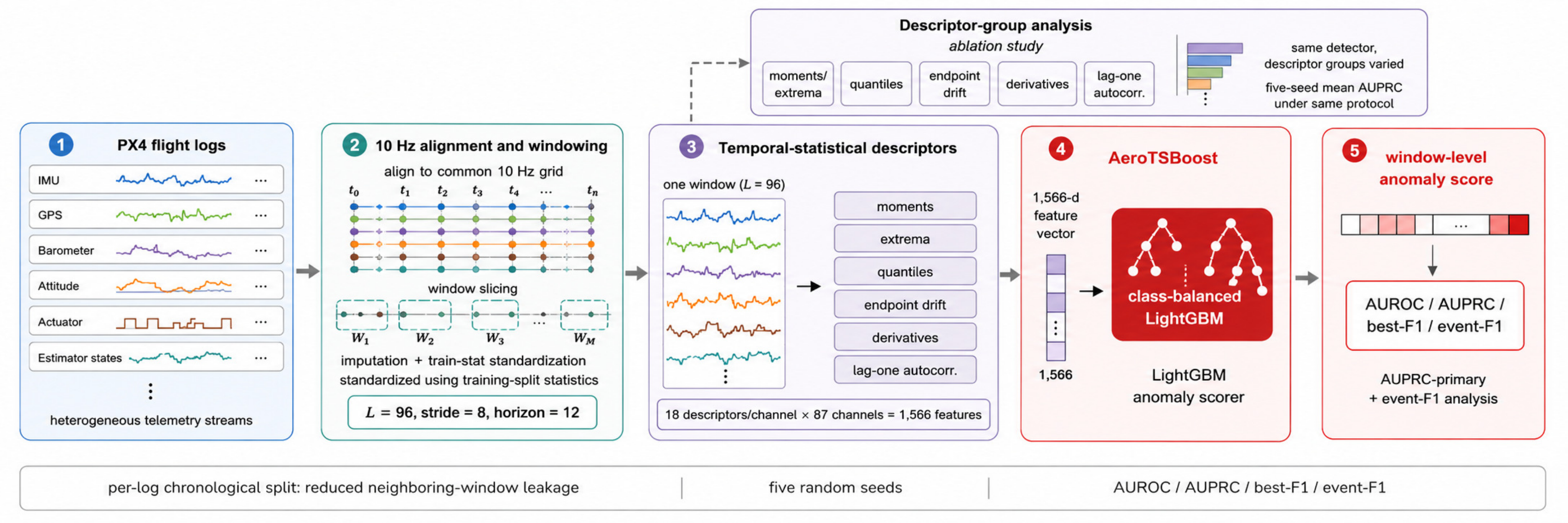}
  \caption{Overview of AeroTSBoost. Heterogeneous PX4 telemetry streams are aligned to a common 10 Hz grid, imputed, standardized using training-split statistics, and sliced into fixed-length windows with length $L=96$, stride $8$, and horizon $12$. Each window is transformed into temporal-statistical descriptors, including moments, extrema, quantiles, endpoint drift, derivatives, and lag-one autocorrelation for 87 retained telemetry channels, yielding a 1,566-dimensional feature vector. AeroTSBoost scores each window with a class-balanced LightGBM detector. The upper branch summarizes descriptor-group ablations under the same detector and protocol, while the bottom strip indicates the chronological, purged, and leave-log-out evaluation protocols, five random seeds, and the AUROC, AUPRC, best-F1, and event-F1 metrics.}
  \label{fig:overview}
\end{figure}

We propose AeroTSBoost, a supervised temporal-statistical boosting framework for UAV state-estimation anomaly mining. The main idea is to convert each aligned multivariate flight window into a compact but expressive set of deterministic temporal descriptors, then train a class-balanced boosted-tree detector. This design emphasizes stability and reproducibility: the representation does not require stochastic neural pretraining, the detector can be trained with standard tabular learners, and every reported metric is generated from saved seed-level outputs. Fig.~\ref{fig:overview} gives an overview of the pipeline.

The contributions are as follows:
\begin{itemize}
  \item We formulate UAV state-estimation anomaly mining as multivariate telemetry window classification and evaluate it with leakage-aware chronological, purged, and leave-log-out protocols plus event-level metrics.
  \item We introduce AeroTSBoost, a temporal-statistical boosting detector that mines distribution, quantile, drift, derivative, and autocorrelation evidence from high-dimensional UAV logs.
  \item We provide a broad empirical comparison against classical detectors, neural autoencoders, tree and linear supervised baselines, GCAD, xLSTMAD, and CATCH adapted to UAV-SEAD.
  \item We report five-seed strict-protocol results, precision-recall analysis, UAV failure-family breakdowns, telemetry-family importance analysis, feature-group ablations, and external ALFA validation on real UAV fault logs.
\end{itemize}

\section{Related Work}
\subsection{Multivariate Time-Series Anomaly Detection}
Classical anomaly detectors remain important baselines for telemetry mining. PCA reconstruction identifies deviations from a low-dimensional subspace~\cite{jolliffe2002principal}, isolation forests isolate sparse abnormal regions~\cite{liu2008isolation}, and local outlier factor estimates density-based deviation~\cite{breunig2000lof}. These methods are simple, reproducible, and often strong when telemetry windows can be represented as tabular feature vectors.

Deep multivariate time-series detectors learn richer temporal representations through reconstruction, forecasting, graph structure, or attention. Graph deviation networks model sensor dependencies for anomaly detection~\cite{deng2021gdn}. Transformer-based approaches such as Anomaly Transformer~\cite{xu2022anomalytransformer} and TranAD~\cite{tuli2022tranad} capture long-range temporal patterns through attention mechanisms. Recent methods further explore recurrent sequence modeling with xLSTM-style architectures~\cite{xlstmad2025}, frequency-domain patching with channel-aware reconstruction as in CATCH~\cite{catch2025}, and Granger-causality-based dependency modeling as in GCAD~\cite{gcad2025}. These methods are valuable comparisons, but many of them are optimized around reconstruction or forecasting proxies rather than sparse labeled event detection.

\subsection{UAV State-Estimation Anomaly Mining}
UAV anomaly detection has been studied through simulated attacks, injected faults, and telemetry corpora. UAV-SEAD~\cite{kabaoglu2026uavsead} expands this setting to real PX4 flight logs with annotated state-estimation anomalies involving mechanical/electrical issues, external-position failures, global-position failures, and altitude-related failures. ALFA~\cite{keipour2021alfa} provides a complementary fixed-wing UAV fault and anomaly dataset with real flight sequences and ground-truth fault timing. This paper focuses on mining real, non-synthetic UAV anomalies from heterogeneous telemetry windows and evaluates detectors under chronological and stricter leakage-aware protocols rather than a random window split.

\section{Problem Formulation}
Let a flight log be an aligned multivariate sequence $X=\{x_t\}_{t=1}^{T}$, where each $x_t\in\mathbb{R}^{d}$ contains $d$ telemetry channels. Given a window length $L$, stride $r$, and labeling horizon $H$, the log is transformed into windows $W_i\in\mathbb{R}^{L\times d}$. Each window receives a binary label $y_i\in\{0,1\}$ indicating whether its window-horizon interval overlaps an annotated anomaly interval. The learning objective is to estimate a scoring function
\begin{equation}
  s(W_i): \mathbb{R}^{L\times d}\rightarrow [0,1],
\end{equation}
such that anomalous windows and anomaly events receive higher scores than normal windows.

The evaluation protocol has two levels. At the point/window level, we report AUROC, AUPRC, and best F1. AUPRC is the primary ranking metric because anomalous windows are sparse. At the event level, consecutive anomalous windows define ground-truth anomaly events, and consecutive predicted anomalous windows define predicted events. A predicted event is counted as a match when it overlaps a ground-truth event. Event F1 summarizes whether a detector raises coherent alarms around anomaly intervals rather than only ranking isolated windows.

To reduce leakage, we avoid random row or random window sampling. Windows are ordered within each flight log and assigned chronologically to train, validation, and test partitions. This protocol preserves the temporal direction of evaluation and reduces duplicate-context leakage relative to random splitting, which would intermix heavily overlapping windows across partitions. For the base chronological protocol, no purge or embargo gap is applied at split boundaries, so we describe that protocol as reducing neighboring-window leakage rather than eliminating all possible overlap effects.

\section{The Proposed AeroTSBoost}
\subsection{Telemetry Alignment and Windowing}
Raw PX4 logs are first converted into aligned 10 Hz multivariate CSV files. For each log, numeric telemetry channels are retained after removing time, label, and anomaly-type columns. Channels that appear in at least 60\% of usable logs are kept, yielding 87 telemetry channels in the final processed UAV-SEAD split. Missing values are repaired deterministically by bidirectional interpolation followed by zero filling for any remaining non-finite entries. The resulting sequence is standardized channel-wise using statistics computed from the training split only.

Given a standardized sequence $X=\{x_t\}_{t=1}^{T}$, AeroTSBoost uses fixed-length sliding windows with length $L=96$, stride 8, and horizon $H=12$. A window starts at index $a$ and contains $W_a=X_{a:a+L-1}\in\mathbb{R}^{L\times d}$. Its label is positive if any annotated anomaly occurs in $X_{a:a+L+H-1}$; otherwise it is negative. This definition allows the detector to score windows that overlap or immediately precede annotated state-estimation failures while preserving a fixed input representation across logs.

\subsection{Temporal-Statistical Descriptor Mining}
For each window $W\in\mathbb{R}^{L\times d}$, AeroTSBoost computes deterministic descriptors independently for each telemetry channel and concatenates them across channels. Let $w^{(c)}=(w^{(c)}_1,\ldots,w^{(c)}_L)$ be channel $c$ in a window, and let $\delta^{(c)}_t=w^{(c)}_{t+1}-w^{(c)}_t$ denote first-order differences. The per-channel descriptor vector is
\begin{align}
\phi_c(W) = [&\mu, \sigma, \min, \max, \max-\min, q_{.10}, q_{.25}, \nonumber\\
&q_{.50}, q_{.75}, q_{.90}, w_1, w_L, w_L-w_1, \nonumber\\
&\mu_\delta, \sigma_\delta, \mu_{|\delta|}, \max|\delta|, \rho_1],
\end{align}
where all statistics are computed over the $L$ samples of channel $c$, $q_p$ is the empirical $p$-quantile, and $\rho_1$ is lag-one autocorrelation computed from adjacent centered values with a small numerical floor in the denominator. The final window representation is
\begin{equation}
  \phi(W)=[\phi_1(W),\ldots,\phi_d(W)]\in\mathbb{R}^{18d}.
\end{equation}
With $d=87$ UAV-SEAD telemetry channels, AeroTSBoost uses 1,566 input features per window.

The descriptor groups are designed to expose different failure signatures in UAV state estimation. Moments, extrema, and ranges capture level shifts and amplitude excursions; quantiles capture heavy-tailed or asymmetric deviations; endpoints and endpoint drift capture gradual estimator displacement; first-order difference summaries capture local volatility and abrupt control-response changes; and lag-one autocorrelation captures changes in short-range temporal dependency. Cross-channel interactions are not hand-coded; they are learned by the boosted detector from the concatenated per-channel descriptors. Because these descriptors are deterministic, the representation can be regenerated without stochastic pretraining or learned feature extraction.

\subsection{Boosted Detector}
\label{subsec:boosted_detector}
AeroTSBoost trains a LightGBM binary classifier~\cite{ke2017lightgbm} on $\phi(W_i)$ with the window labels described above. The detector uses 1,600 trees as an upper bound, learning rate 0.025, 64 leaves, minimum child samples 80, subsampling ratio 0.9, column sampling ratio 0.75, $\ell_1$ regularization 0.2, $\ell_2$ regularization 8.0, balanced class weights, and validation early stopping with patience 80 using average precision. The anomaly score $s(W_i)$ is the positive-class output returned by the fitted ensemble; it is used as a ranking score rather than as a calibrated posterior probability.

LightGBM is well suited to this setting because it can exploit sparse nonlinear interactions among telemetry descriptors while remaining efficient on hundreds of thousands of windows. Computationally, descriptor extraction costs $O(NLd)$ time for $N$ windows, window length $L$, and $d$ telemetry channels, and stores $O(N\cdot18d)$ tabular features. For fixed $L=96$, this scales linearly with the number of windows and telemetry channels; inference through the boosted detector is linear in the number of trees and tree depth per window. Validation scores are used for model selection and early stopping. AUROC and AUPRC are threshold independent, while best F1 and event-F1 are diagnostic threshold-swept metrics reported with the same best-F1 thresholding rule for all methods.

\section{Experiments}
\subsection{Dataset}
We evaluate on UAV-SEAD~\cite{kabaoglu2026uavsead}, a real-world UAV state-estimation anomaly dataset containing PX4 flight logs and anomaly annotations. The logs contain heterogeneous state-estimation telemetry collected from operational UAV flights, including inertial, position, attitude, actuator, and estimator-related streams. We use only logs with usable anomaly annotations, align all retained telemetry streams to 10 Hz, and keep the numeric channels that satisfy the coverage rule described in Section~IV.

\begin{figure}[t]
  \centering
  \includegraphics[width=\linewidth]{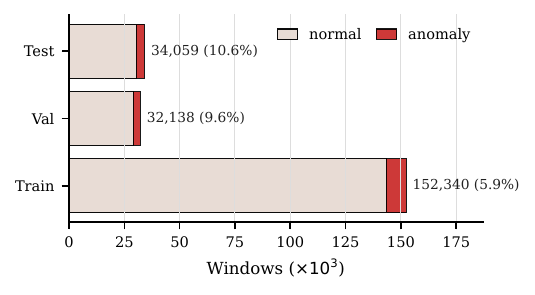}
  \caption{Window-level composition of the per-log chronological UAV-SEAD split. Each bar shows normal and anomalous windows in the train, validation, and test partitions; labels report the total number of windows and anomaly rate.}
  \label{fig:dataset_protocol}
\end{figure}

The processed benchmark used in this study contains 1,389 usable annotated logs, 1,892,063 aligned rows, and 87 telemetry channels. Sliding-window construction with $L=96$, stride 8, and horizon $H=12$ produces 218,537 windows. The per-log chronological split contains 152,340 training windows, 32,138 validation windows, and 34,059 test windows. Table~\ref{tab:dataset_stats} summarizes the preprocessing and split statistics, and Fig.~\ref{fig:dataset_protocol} visualizes the split composition and anomaly rate.

\begin{table}[t]
\caption{UAV-SEAD preprocessing and split statistics used in this study. Window counts are computed after 10 Hz alignment, interpolation, sliding-window construction, and per-log chronological splitting.}
\label{tab:dataset_stats}
\centering
\footnotesize
\setlength{\tabcolsep}{3.5pt}
\begin{tabular}{lc}
\toprule
Item & Value \\
\midrule
Usable annotated logs & 1,389 \\
Aligned rows at 10 Hz & 1,892,063 \\
Telemetry channels & 87 \\
Window length / stride / horizon & 96 / 8 / 12 \\
AeroTSBoost descriptor dimension & 1,566 \\
Total windows & 218,537 \\
Anomalous windows & 15,694 (7.18\%) \\
Training windows & 152,340 (5.92\%) \\
Validation windows & 32,138 (9.59\%) \\
Test windows & 34,059 (10.58\%) \\
Split policy & Per-log chronological \\
\bottomrule
\end{tabular}
\end{table}

For related UAV telemetry external validation, we additionally evaluate on ALFA~\cite{keipour2021alfa}, a real fixed-wing UAV fault and anomaly dataset. We use the processed CSV release, exclude the sequence without ground truth, align numeric telemetry streams to 10 Hz, and convert failure-status annotations into binary window labels with the same window length, stride, and horizon. This produces 46 sequences, 47,138 aligned rows, 443 numeric channels, 5,298 windows, and 968 anomalous windows. With 443 channels, the same 18 descriptors per channel give 7,974 features per ALFA window. Because ALFA is smaller than UAV-SEAD and targets different fault mechanisms, we use it as an external validation set rather than as the primary benchmark, reporting five-seed leave-sequence-out results with the same AeroTSBoost hyperparameters and the same core classical/supervised comparison set.

\subsection{Baselines}
We compare AeroTSBoost against three baseline families under the same window construction and the corresponding evaluation protocol where reported. Classical anomaly detectors include PCA reconstruction~\cite{jolliffe2002principal}, Isolation Forest~\cite{liu2008isolation}, and LOF~\cite{breunig2000lof}. These methods are fitted on training-window statistics and score test windows by reconstruction error, isolation depth, or local density deviation. Neural and supervised baselines include an LSTM autoencoder (LSTM-AE) for sequence reconstruction~\cite{malhotra2016lstm}, RandomForest~\cite{breiman2001random}, and an elastic-net logistic classifier trained with stochastic gradient descent (Linear-SGD)~\cite{bottou2010large,pedregosa2011scikit}. RandomForest and Linear-SGD use the same temporal-statistical feature matrix as AeroTSBoost, isolating the effect of the boosted detector from the feature representation. Recent deep or structured baselines include GCAD~\cite{gcad2025}, xLSTMAD~\cite{xlstmad2025}, and CATCH~\cite{catch2025}, representing Granger-causality-based dependency modeling, recurrent sequence modeling, and frequency patching, respectively. LOF is included in the broad chronological comparison; the stricter protocol table focuses on scalable classical detectors, supervised tabular baselines, and the recent sequence/structured baselines evaluated over five seeds.

\subsection{Metrics, Protocol, and Hardware}
All main results are reported over five random seeds. We use AUROC, AUPRC, best F1, and event F1. AUPRC is treated as the primary metric because the positive class is rare and because high precision at useful recall is critical for anomaly mining. Best F1 is the maximum point-wise F1 over score thresholds on the test labels, and event F1 uses the corresponding threshold to measure overlap between predicted and annotated anomaly segments. Thus, best F1 and event F1 are diagnostic upper-bound operating-point summaries rather than validation-frozen deployment metrics; AUPRC remains the primary threshold-independent comparison. Tables and figures are generated from saved seed-level metric and score files to avoid manual transcription.

In addition to the base per-log chronological protocol, we evaluate two stricter checks for leakage and generalization. The purged chronological protocol discards windows within an embargo interval of at least $L+H$ samples around train/validation and validation/test boundaries inside each log, eliminating overlapping window-horizon context across adjacent split boundaries. The leave-log-out protocol splits whole flight logs into train, validation, and test partitions, so test windows come from logs unseen during training.

Experiments were run on a Linux workstation with an Intel Xeon E5-2698 v4 CPU (20 cores, 40 threads), 251 GiB RAM, and four NVIDIA Tesla V100-DGXS GPUs with 32 GiB memory each (driver 535.230.02). The environment used Python 3.11.15, PyTorch 2.4.0+cu121, LightGBM 4.6.0, and scikit-learn 1.8.0. GPU-capable deep baselines were executed on GPU when supported; classical, linear, and tree baselines used CPU execution with parallel workers where available.

\section{Main Results}
We first evaluate AeroTSBoost under stricter split protocols because they directly address neighboring-window overlap and log-specific memorization. Table~\ref{tab:robustness_protocols} reports the strict-protocol comparison set over five seeds, including classical detectors, supervised tabular baselines, neural reconstruction, recurrent sequence modeling, Granger-causality-based dependency modeling, and frequency-domain patching. Under the purged chronological protocol, AeroTSBoost obtains the best AUPRC ($0.6066\pm0.0193$), improving over Moments-only by 2.83 absolute points and over RandomForest by 5.88 absolute points. The recent deep or structured baselines are substantially lower under this purged protocol, with CATCH, LSTM-AE, xLSTMAD, and GCAD reaching AUPRC values between $0.1108$ and $0.1473$. Under leave-log-out evaluation, where all test windows come from held-out flight logs, AeroTSBoost again reaches the highest AUPRC ($0.6388\pm0.0315$), ahead of Moments-only ($0.6285\pm0.0274$) and RandomForest ($0.6122\pm0.0449$). Linear-SGD, Isolation Forest, PCA, CATCH, LSTM-AE, xLSTMAD, and GCAD remain far below the three supervised tree-based/descriptor methods, indicating that the strict-protocol result is not explained only by linear separability, generic unsupervised scoring, or self-supervised sequence objectives.

\begin{table}[t]
\caption{Strict-protocol anomaly detection results for classical, supervised tabular, neural, recurrent, Granger-causality-based, and frequency-domain baselines. Values are mean $\pm$ standard deviation over five seeds. Higher is better for all metrics; bold marks the best result within each protocol.}
\label{tab:robustness_protocols}
\centering
\scriptsize
\setlength{\tabcolsep}{2.2pt}
\begin{tabular}{llcccc}
\toprule
Protocol & Method & AUROC $\uparrow$ & AUPRC $\uparrow$ & Best F1 $\uparrow$ & Event F1 $\uparrow$ \\
\midrule
Purged chronological & AeroTSBoost & $\boldsymbol{0.9436\pm0.0029}$ & $\boldsymbol{0.6066\pm0.0193}$ & $\boldsymbol{0.6093\pm0.0107}$ & $0.4774\pm0.0237$ \\
Purged chronological & Moments-only & $0.9369\pm0.0032$ & $0.5783\pm0.0080$ & $0.6007\pm0.0068$ & $\boldsymbol{0.4903\pm0.0173}$ \\
Purged chronological & RandomForest & $0.9340\pm0.0019$ & $0.5479\pm0.0043$ & $0.5514\pm0.0061$ & $0.4169\pm0.0245$ \\
Purged chronological & CATCH & $0.7400\pm0.0013$ & $0.1473\pm0.0005$ & $0.2496\pm0.0018$ & $0.1833\pm0.0035$ \\
Purged chronological & PCA & $0.7059\pm0.0012$ & $0.1379\pm0.0004$ & $0.2260\pm0.0008$ & $0.1649\pm0.0053$ \\
Purged chronological & xLSTMAD & $0.6918\pm0.0103$ & $0.1416\pm0.0046$ & $0.2236\pm0.0100$ & $0.2525\pm0.0568$ \\
Purged chronological & Linear-SGD & $0.7990\pm0.0020$ & $0.1826\pm0.0017$ & $0.3235\pm0.0029$ & $0.1605\pm0.0031$ \\
Purged chronological & LSTM-AE & $0.7355\pm0.0017$ & $0.1453\pm0.0020$ & $0.2488\pm0.0022$ & $0.1896\pm0.0101$ \\
Purged chronological & Isolation Forest & $0.6927\pm0.0051$ & $0.1577\pm0.0013$ & $0.2335\pm0.0054$ & $0.2602\pm0.0246$ \\
Purged chronological & GCAD & $0.6942\pm0.0021$ & $0.1108\pm0.0008$ & $0.2255\pm0.0010$ & $0.1487\pm0.0030$ \\
Leave-log-out & AeroTSBoost & $\boldsymbol{0.9330\pm0.0071}$ & $\boldsymbol{0.6388\pm0.0315}$ & $\boldsymbol{0.5989\pm0.0264}$ & $0.4108\pm0.0186$ \\
Leave-log-out & Moments-only & $0.9286\pm0.0058$ & $0.6285\pm0.0274$ & $0.5881\pm0.0190$ & $\boldsymbol{0.4111\pm0.0242}$ \\
Leave-log-out & RandomForest & $0.9299\pm0.0097$ & $0.6122\pm0.0449$ & $0.5843\pm0.0211$ & $0.3921\pm0.0579$ \\
Leave-log-out & CATCH & $0.7471\pm0.0555$ & $0.2115\pm0.0435$ & $0.2968\pm0.0335$ & $0.1890\pm0.0212$ \\
Leave-log-out & PCA & $0.7163\pm0.0604$ & $0.1905\pm0.0416$ & $0.2808\pm0.0403$ & $0.1262\pm0.0201$ \\
Leave-log-out & xLSTMAD & $0.6911\pm0.0721$ & $0.1878\pm0.0494$ & $0.2840\pm0.0373$ & $0.2842\pm0.0246$ \\
Leave-log-out & Linear-SGD & $0.8124\pm0.0368$ & $0.2411\pm0.0534$ & $0.3952\pm0.0706$ & $0.1322\pm0.0145$ \\
Leave-log-out & LSTM-AE & $0.7378\pm0.0627$ & $0.2196\pm0.0634$ & $0.3077\pm0.0400$ & $0.2496\pm0.0169$ \\
Leave-log-out & Isolation Forest & $0.6896\pm0.0693$ & $0.2097\pm0.0520$ & $0.2868\pm0.0409$ & $0.2228\pm0.0128$ \\
Leave-log-out & GCAD & $0.7328\pm0.0526$ & $0.1826\pm0.0248$ & $0.2684\pm0.0296$ & $0.0975\pm0.0240$ \\
\bottomrule
\end{tabular}
\end{table}

\begin{figure}[t]
  \centering
  \includegraphics[width=\linewidth]{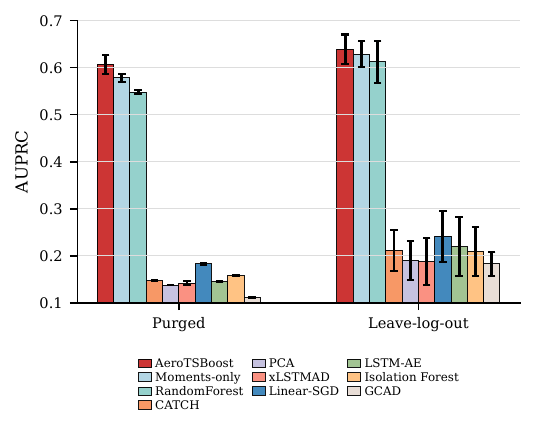}
  \caption{AUPRC under stricter split protocols for the strict-protocol comparison set. Bars show mean AUPRC over five seeds; error bars denote one standard deviation.}
  \label{fig:robustness_auprc}
\end{figure}

Fig.~\ref{fig:robustness_auprc} visualizes the strict-protocol AUPRC comparison. The purged split is substantially harder than the main chronological split because many boundary-adjacent windows are discarded and the number of positive test windows drops from 3,602 to 1,248. The leave-log-out split is also harder because training and test windows come from disjoint flight logs. AeroTSBoost remains the best ranking method in both settings after adding the deep and structured baselines to the same strict protocols. At the same time, Moments-only obtains slightly higher diagnostic event-F1 in the strict protocols, showing that thresholded event alarms can favor simple low-order summaries even when the full temporal-statistical descriptor set improves ranking.

For completeness, Table~\ref{tab:main_results} reports the broader chronological comparison, including recent neural, recurrent, Granger-causality-based, and frequency-domain baselines. AeroTSBoost reaches $0.7516\pm0.0043$ AUPRC, $0.9516\pm0.0009$ AUROC, $0.6860\pm0.0043$ diagnostic best F1, and $0.5342\pm0.0108$ diagnostic event F1. The strongest non-AeroTSBoost baseline in this broader comparison is RandomForest, with $0.6937\pm0.0047$ AUPRC and $0.4801\pm0.0168$ event F1. AeroTSBoost therefore improves chronological AUPRC by 5.79 absolute points over RandomForest and by a larger margin over CATCH, xLSTMAD, GCAD, LSTM-AE, PCA, Isolation Forest, LOF, and Linear-SGD.

\begin{table}[t]
\caption{Chronological UAV-SEAD anomaly detection results over five seeds. Values are mean $\pm$ standard deviation. Higher is better for all metrics; bold marks the best result in each column. Best F1 and Event F1 are diagnostic test-label threshold-swept summaries.}
\label{tab:main_results}
\centering
\footnotesize
\setlength{\tabcolsep}{3.5pt}
\begin{tabular}{lcccc}
\toprule
Method & AUROC $\uparrow$ & AUPRC $\uparrow$ & Best F1 $\uparrow$ & Event F1 $\uparrow$ \\
\midrule
AeroTSBoost & $\boldsymbol{0.9516\pm0.0009}$ & $\boldsymbol{0.7516\pm0.0043}$ & $\boldsymbol{0.6860\pm0.0043}$ & $\boldsymbol{0.5342\pm0.0108}$ \\
CATCH & $0.7465\pm0.0010$ & $0.2427\pm0.0010$ & $0.3151\pm0.0011$ & $0.2245\pm0.0053$ \\
PCA & $0.7091\pm0.0008$ & $0.2095\pm0.0006$ & $0.2980\pm0.0009$ & $0.2039\pm0.0021$ \\
RandomForest & $0.9410\pm0.0018$ & $0.6937\pm0.0047$ & $0.6309\pm0.0059$ & $0.4801\pm0.0168$ \\
xLSTMAD & $0.7057\pm0.0044$ & $0.2271\pm0.0029$ & $0.2990\pm0.0036$ & $0.2609\pm0.0187$ \\
LOF & $0.6473\pm0.0017$ & $0.1594\pm0.0016$ & $0.2552\pm0.0019$ & $0.1690\pm0.0029$ \\
LSTM-AE & $0.7480\pm0.0018$ & $0.2421\pm0.0017$ & $0.3231\pm0.0026$ & $0.2269\pm0.0021$ \\
Isolation Forest & $0.7200\pm0.0030$ & $0.2483\pm0.0052$ & $0.3163\pm0.0038$ & $0.2281\pm0.0104$ \\
GCAD & $0.7037\pm0.0018$ & $0.1768\pm0.0047$ & $0.2986\pm0.0020$ & $0.1774\pm0.0060$ \\
Linear-SGD & $0.8015\pm0.0010$ & $0.2745\pm0.0020$ & $0.4438\pm0.0027$ & $0.2193\pm0.0043$ \\
\bottomrule
\end{tabular}
\end{table}

\begin{figure}[t]
  \centering
  \includegraphics[width=\linewidth]{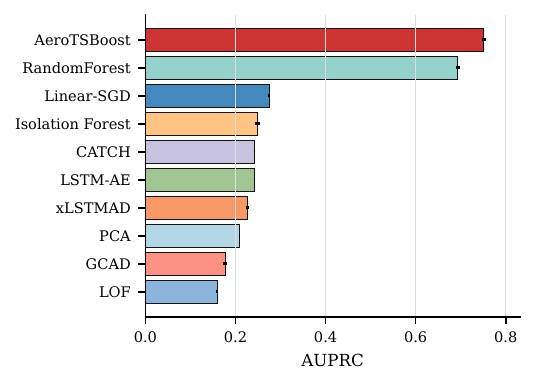}
  \caption{Chronological AUPRC comparison on UAV-SEAD. Bars show mean AUPRC over five seeds under the per-log chronological split; error bars denote one standard deviation.}
  \label{fig:main_auprc}
\end{figure}

Fig.~\ref{fig:main_auprc} visualizes the chronological AUPRC ranking across the reported comparison set. AeroTSBoost achieves the best AUPRC, while RandomForest is the strongest non-AeroTSBoost baseline, suggesting that labeled temporal-statistical evidence is especially effective for this dataset. In contrast, unsupervised reconstruction, local density, recurrent, Granger-causality-based, and frequency-domain methods produce substantially lower AUPRC under the same split and evaluation.

\begin{figure}[t]
  \centering
  \includegraphics[width=\linewidth]{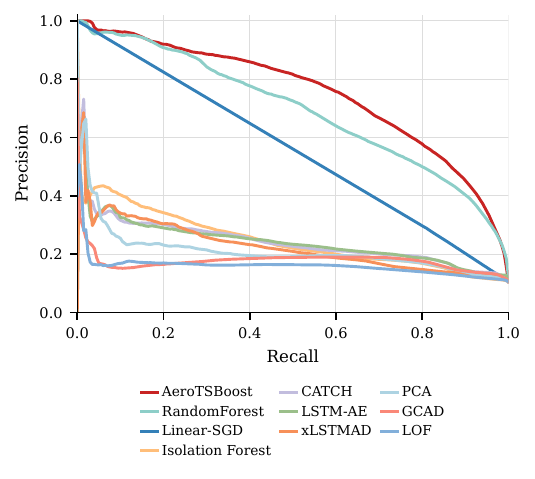}
  \caption{Precision-recall curves on UAV-SEAD. Curves are averaged over five seeds for the reported methods under the same per-log chronological evaluation protocol.}
  \label{fig:pr}
\end{figure}

The averaged precision-recall curves in Fig.~\ref{fig:pr} provide a threshold-independent view of the chronological ranking behavior. AeroTSBoost maintains higher precision over a broad recall range, which is consistent with its superior AUPRC. This is operationally important because UAV anomaly mining typically requires analysts or downstream safety monitors to prioritize a small number of high-confidence anomalous intervals.

Table~\ref{tab:alfa_external} reports the external ALFA leave-sequence-out validation. AeroTSBoost obtains the best AUROC, AUPRC, and diagnostic best F1, with $0.9259\pm0.0076$ AUPRC compared with $0.8835\pm0.0797$ for RandomForest and $0.8700\pm0.0481$ for moments-only. RandomForest obtains the highest event-F1 on ALFA, so the external result should be read as stronger evidence for ranking robustness than for uniform dominance at every thresholded operating point.

\begin{table}[t]
\caption{External ALFA leave-sequence-out fault-detection results over five seeds. Values are mean $\pm$ standard deviation. Higher is better for all metrics; bold marks the best result in each column.}
\label{tab:alfa_external}
\centering
\scriptsize
\setlength{\tabcolsep}{2.7pt}
\begin{tabular}{lcccc}
\toprule
Method & AUROC $\uparrow$ & AUPRC $\uparrow$ & Best F1 $\uparrow$ & Event F1 $\uparrow$ \\
\midrule
AeroTSBoost & $\boldsymbol{0.9766\pm0.0074}$ & $\boldsymbol{0.9259\pm0.0076}$ & $\boldsymbol{0.8736\pm0.0193}$ & $0.7692\pm0.1107$ \\
RandomForest & $0.9621\pm0.0298$ & $0.8835\pm0.0797$ & $0.8123\pm0.1048$ & $\boldsymbol{0.8410\pm0.1041}$ \\
Moments-only & $0.9541\pm0.0321$ & $0.8700\pm0.0481$ & $0.8132\pm0.0553$ & $0.6492\pm0.1427$ \\
Linear-SGD & $0.8098\pm0.1215$ & $0.4549\pm0.2273$ & $0.5973\pm0.2209$ & $0.5029\pm0.0682$ \\
Isolation Forest & $0.6496\pm0.1161$ & $0.2627\pm0.1040$ & $0.3744\pm0.0753$ & $0.3704\pm0.0979$ \\
PCA & $0.6015\pm0.1495$ & $0.2197\pm0.1278$ & $0.3345\pm0.0769$ & $0.3387\pm0.1121$ \\
\bottomrule
\end{tabular}
\end{table}

\section{Event-Level and UAV-Specific Analysis}
\begin{figure}[t]
  \centering
  \includegraphics[width=0.98\textwidth]{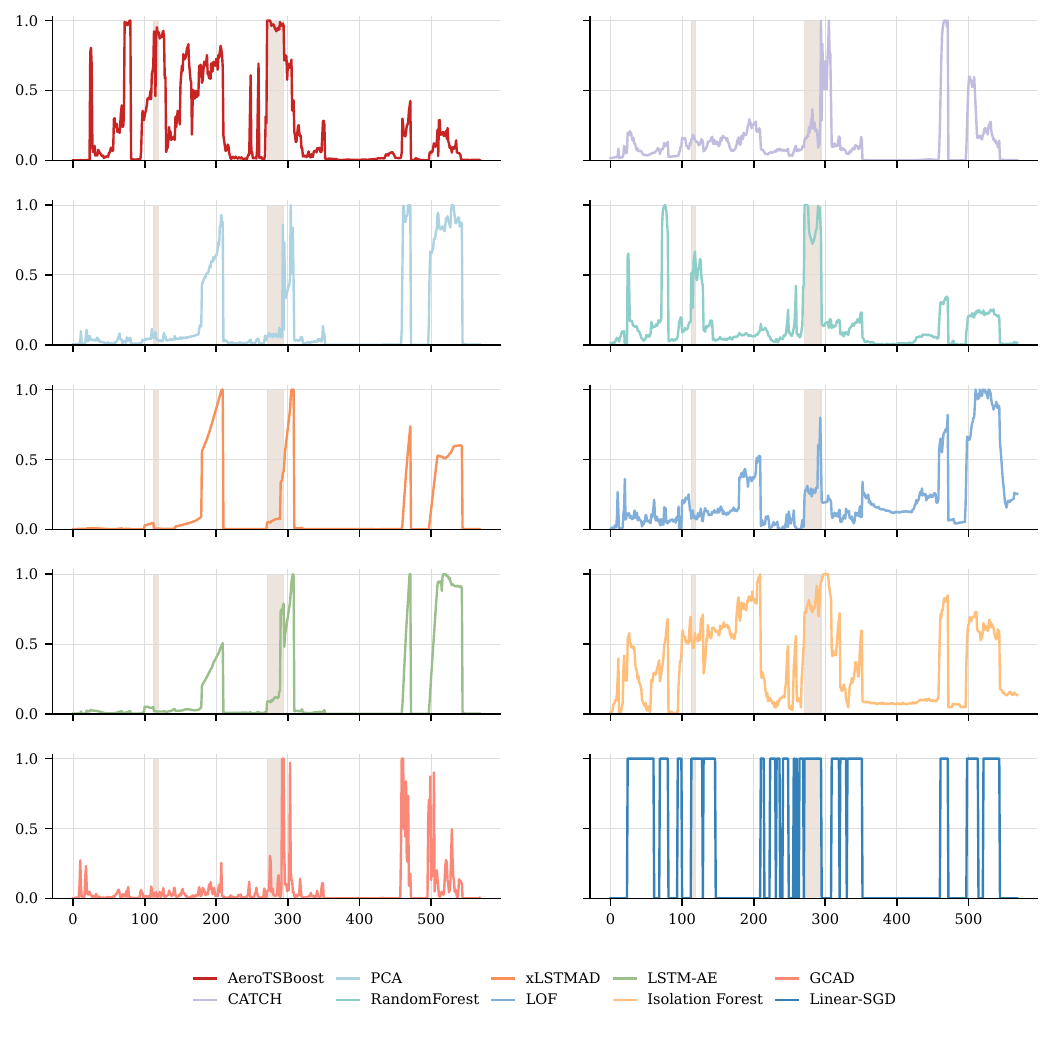}
  \caption{First held-out chronological test anomaly event. Each panel shows one detector's score normalized within the displayed window; shaded intervals denote ground-truth anomalous windows and the legend maps panels to methods.}
  \label{fig:case}
\end{figure}

Window-level ranking alone does not fully describe whether a detector produces useful alarms around real anomaly events. To avoid manual case selection, Fig.~\ref{fig:case} shows the first held-out anomaly event encountered in the chronological test split. Scores are normalized within the displayed window for visualization only, and the shaded region marks the annotated anomaly interval.

AeroTSBoost produces a sharper and more event-aligned response than the baselines in this example. The score rises near the annotated event and remains comparatively controlled outside the event region. This behavior helps explain the event-F1 advantage in Table~\ref{tab:main_results}. The result also illustrates why event-level evaluation is needed: methods with similar AUROC can differ substantially in how coherently they raise alarms around contiguous anomaly intervals.

\begin{figure}[t]
  \centering
  \includegraphics[width=\linewidth]{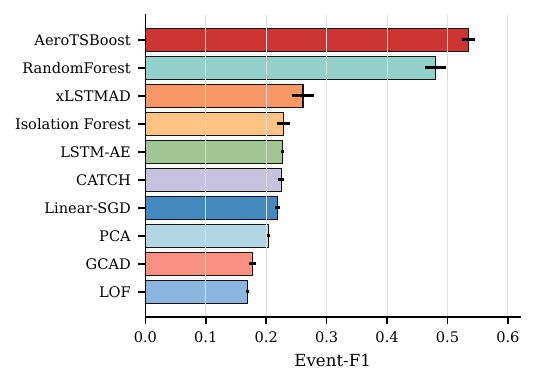}
  \caption{Event-level F1 comparison on UAV-SEAD. Bars show mean event-F1 over five seeds under the same test-label threshold-sweep rule; error bars denote one standard deviation.}
  \label{fig:event_f1}
\end{figure}

Fig.~\ref{fig:event_f1} further compares event-F1 across leading methods. AeroTSBoost remains the strongest method, followed by RandomForest and xLSTMAD among the reported baselines. The consistency between AUPRC and event-F1 indicates that the proposed representation improves both ranking quality and event-level alarm coherence under the main comparison protocol.

\begin{table}[t]
\caption{AeroTSBoost performance by UAV-SEAD anomaly family. Values are mean $\pm$ standard deviation over five seeds. Chron. denotes the per-log chronological protocol; LLO denotes leave-log-out. Count columns report positive windows/anomaly events, with LLO counts averaged over seeds.}
\label{tab:fault_type_breakdown}
\centering
\scriptsize
\setlength{\tabcolsep}{2.2pt}
\begin{tabular}{lcccccc}
\toprule
Failure family & Chron. pos./evt. & LLO pos./evt. & Chron. AUPRC $\uparrow$ & Chron. Event F1 $\uparrow$ & LLO AUPRC $\uparrow$ & LLO Event F1 $\uparrow$ \\
\midrule
External Position & $1,707/100$ & $1,090/29$ & $0.7751\pm0.0038$ & $0.5604\pm0.0106$ & $0.6596\pm0.0745$ & $0.4380\pm0.0348$ \\
Global Position & $641/26$ & $585/10$ & $0.3381\pm0.0066$ & $0.2105\pm0.0154$ & $0.2576\pm0.2117$ & $0.1063\pm0.0454$ \\
Altitude & $571/45$ & $580/15$ & $0.1835\pm0.0062$ & $0.2361\pm0.0211$ & $0.1011\pm0.0149$ & $0.1143\pm0.0411$ \\
Mechanical/Electrical & $683/30$ & $227/6$ & $0.2112\pm0.0095$ & $0.1804\pm0.0233$ & $0.4093\pm0.0838$ & $0.2884\pm0.1050$ \\
\bottomrule
\end{tabular}
\end{table}

Table~\ref{tab:fault_type_breakdown} breaks down AeroTSBoost performance by UAV-SEAD anomaly family and reports the number of positive windows and anomaly events used for each family. External-position failures are the most detectable family, reaching $0.7751\pm0.0038$ AUPRC under the chronological protocol and $0.6596\pm0.0745$ AUPRC under leave-log-out evaluation. Global-position, altitude, and mechanical/electrical failures are more difficult and show larger cross-log variability, reflecting differences in duration, telemetry footprint, and log distribution across failure types. This analysis supports the UAV-specific interpretation of the task: the benchmark is not a homogeneous anomaly label, but a mixture of state-estimation failure mechanisms with different observable signatures.

\begin{figure}[t]
  \centering
  \includegraphics[width=\linewidth]{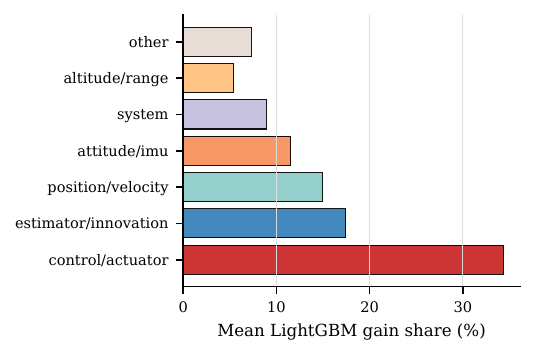}
  \caption{Telemetry-family importance for AeroTSBoost. Bars show LightGBM gain share averaged over the chronological, purged, and leave-log-out protocols after normalizing gains within each protocol.}
  \label{fig:telemetry_family_importance}
\end{figure}

Fig.~\ref{fig:telemetry_family_importance} summarizes the telemetry families most used by the boosted detector. Control/actuator features account for the largest mean gain share (34.42\%), followed by estimator/innovation (17.45\%) and position/velocity features (15.00\%). The resulting importance profile is consistent with the state-estimation setting: detectable failures often involve commanded-response mismatch, estimator residual behavior, and position or velocity inconsistencies rather than a single isolated sensor stream.

\section{Ablation Study}
\begin{figure}[t]
  \centering
  \includegraphics[width=\linewidth]{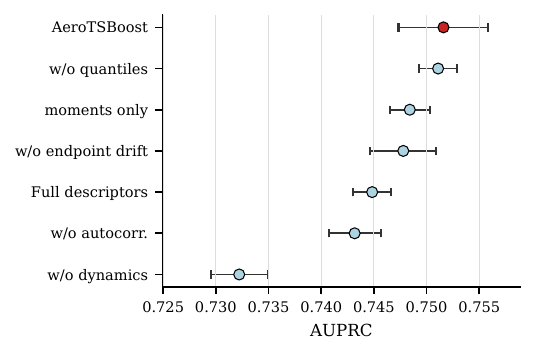}
  \caption{AeroTSBoost descriptor-group ablation. Points show mean AUPRC over five seeds under the same LightGBM protocol; horizontal error bars denote one standard deviation.}
  \label{fig:feature_ablation}
\end{figure}

Fig.~\ref{fig:feature_ablation} evaluates the sensitivity of AeroTSBoost to temporal-statistical descriptor groups under the same LightGBM training and evaluation protocol. The goal is not to introduce additional model variants, but to test whether the proposed descriptor representation is supported by multiple complementary signal families. Each ablation keeps the data split, classifier family, and metric computation fixed, and changes only the descriptor groups available to the detector.

The results show that temporal dynamics produce the largest observed drop among the tested removals. AeroTSBoost obtains $0.7516$ AUPRC, whereas removing dynamic descriptors reduces AUPRC to $0.7322$, a drop of 1.94 absolute points. Removing lag-one autocorrelation also degrades performance to $0.7432$. The moments-only variant remains competitive at $0.7484$, indicating that robust low-order statistics provide a strong base, while temporal descriptors add complementary ranking evidence. A broader descriptor set reaches $0.7448$, suggesting that simply adding more summaries does not necessarily improve sparse anomaly ranking.

These patterns are consistent with the structure of real UAV telemetry anomalies. State-estimation failures often appear as short-term drifts, volatility changes, delayed sensor response, or cross-channel inconsistency rather than as isolated large-amplitude deviations. Dynamic, quantile, drift, and autocorrelation descriptors therefore provide complementary evidence for the boosted detector. The ablation results support the main design choice of AeroTSBoost, while also showing that the advantage comes from incremental temporal evidence on top of already strong statistical summaries.

\section{Discussion}
The results suggest that real UAV state-estimation anomaly mining benefits from explicit temporal-statistical evidence. Many UAV-SEAD anomalies manifest as multi-channel shifts, drifts, volatility changes, or short-term temporal dependency changes. AeroTSBoost represents these behaviors directly through deterministic descriptors and then uses boosting to learn nonlinear interactions among channels and descriptor groups. This is different from reconstruction or forecasting methods, which may assign low anomaly scores to failures that are easy to reconstruct or forecast under the learned proxy objective.

The comparison with recent sequence baselines should not be interpreted as a general claim that deep models are weak. Instead, it highlights a mismatch between common self-supervised anomaly objectives and sparse labeled event detection in this particular UAV telemetry setting. When labels are available, supervised boosting over well-designed temporal descriptors can exploit the sparse boundary directly and with lower training variance.

The stricter protocol checks clarify what the current evidence supports. AeroTSBoost keeps the highest AUPRC under chronological, purged chronological, and leave-log-out evaluation, including when recent neural, recurrent, Granger-causality-based, and frequency-domain baselines are evaluated under the same strict splits. Thus, the main ranking result is not solely an artifact of random-window mixing or immediate boundary overlap. However, moments-only descriptors sometimes yield slightly better threshold-swept event-F1, showing that low-order statistical summaries are already strong alarm features and that temporal descriptors primarily improve ranking robustness rather than uniformly improving every thresholded operating point.

There are also limitations. AeroTSBoost currently depends on labeled anomaly intervals, while many deployment settings may have incomplete or delayed labels. The main empirical study focuses on UAV-SEAD, with ALFA used as a smaller related UAV telemetry external validation set; additional UAV platforms, vehicles, and operating conditions would further strengthen external validity. The current bidirectional imputation is designed for offline log mining; causal imputation would be required for fully online deployment. Finally, the current experiments evaluate offline window-level ranking and threshold-swept operating points. Online deployment would require further analysis of latency, memory footprint, validation-based threshold adaptation, and delayed alarm handling.

\section{Conclusion}
This paper introduced AeroTSBoost for mining UAV state-estimation anomalies from real PX4 telemetry. AeroTSBoost converts aligned flight windows into deterministic temporal-statistical descriptors and uses a class-balanced LightGBM detector to learn nonlinear anomaly evidence across channels and descriptor groups. On UAV-SEAD, it achieves the best five-seed AUPRC among the evaluated classical, supervised tabular, neural reconstruction, recurrent, Granger-causality-based, and frequency-domain baselines. Purged chronological and leave-log-out checks show that the AUPRC advantage persists under stricter split protocols, while ALFA external validation provides additional related UAV telemetry evidence on fixed-wing fault logs. Anomaly-family and telemetry-family analyses connect the detector to UAV state-estimation failure structure. Future work will extend AeroTSBoost toward online deployment, anomaly-type diagnosis, and cross-platform UAV telemetry evaluation.

\section*{Acknowledgements}

The supports provided by Macao Polytechnic University (RP/FCA-01/2025) and Macao Science and Technology Development Fund (FDCT-MOST: 0018/2025/AMJ) enabled us to conduct data collection, analysis, and interpretation, as well as cover expenses related to research materials and participant recruitment. MPU and FDCT investment in our work have significantly contributed to the quality and impact of our research findings.


\end{document}